\documentclass[journal]{IEEEtran}

\usepackage{graphicx} 
\usepackage{amsmath,amssymb,amsfonts}
\usepackage{algorithmic}
\usepackage{graphicx}
\usepackage{subfigure}
\usepackage{textcomp}
\usepackage{makecell}
\usepackage{threeparttable}

\begin{document}

\title{A Thermodynamic based and Data Driven Hybrid Network for Gas Turbine Modeling}
\author{Likun Ren, Haiqin Qin, Kejun Xu
\thanks{All authors are with Qingdao Campus,  Naval Aviation University, Qingdao CO 266041, China
	e-mail: renlikun1988@foxmail.com}}
\maketitle

\begin{abstract}
 The on-wing engine performance is difficult to track for  thermodynamic models because of its inaccurate component maps, and also difficult for data driven methods for their over-fitting to measurement errors. So, we propose a thermodynamic based and data driven hybrid network for gas turbine modeling. Different from thermodynamic models, our network reconstructs the component characteristics in a data-driven way to take component degeneration and individual difference into consideration. Moreover, different from data driven methods, in the training phase, physical based equations and the analytical mathematical description are used to ensure that the optimization converges to the gas turbine's dynamics. A huge number of relaxed quasi steady state flight data to 26970 is used to train and test our hybrid network. The result shows that the accuracy of our hybrid network can reach about 7\% measured by max T6 relative error, 5\% better than map fitting based thermodynamic model and 8\% better than pure data driven method with similar model volume.
\end{abstract}


\section{Introduction}
\label{sec:introduction}
Gas turbine performance modeling has been widely used in the development cycle \cite{lytle1999numerical, sethi2013map}. However, it is rarely applied to the on-wing engine performance evaluation field due to the relatively low accuracy. We think the reasons for the low accuracy mainly lie in the following two points.

First, it is difficult to access accurate component characteristic maps. What is more, the maps are different from engine to engine due to manufacture and assembly tolerance.

Second, the degeneration process of components is hard to model. So it is difficult for performance models to track the gas turbine degeneration trajectory.

Large amount of flight data is the main advantage in on-wing gas turbine performance evaluation field over development cycle. The component characteristics and degeneration trend hidden behind the flight data may be extracted using proper data driven methods \cite{asgari2015gas} \cite{chiras2001nonlinear} \cite{wei2020robust}. However, only data driven methods cannot evaluate gas turbine performance from a comprehensive perspective. One is because data driven models can only cover a few parameters recorded in flight data such as rotational speed and EGT. Other parameters key to performance evaluation can not be calculated using data driven methods. Moreover, data driven methods are easy to run into over-fitting under one single constrained optimization goal.

To use flight data more effectively, many researches focus on component maps fitting. Map scaling and curve fitting are first introduced. Reference \cite{kong2003new} proposed a polynomial scaling method to correct the component maps. In \cite{Kong2006Component} and \cite{kong2007study}, cubic fitting was used to directly fit mass flow and isentropic efficiency. In \cite{li2011nonlinear} and \cite{li2012improved}, scale factors were regressed using quadratic curve.  In \cite{tsoutsanis2014component}, elliptical curves fitting was used to tune component maps. In \cite{tsoutsanis2015transient}, scale factors of different components were regressed by various functions according to the shape of original maps. In \cite{kim2018adaptation}, the overall performance parameters and the local performance parameters were separately adjusted in two steps by scale factors. These scaling and curve fitting methods can correct component maps toward the operating space of certain engine especially in design point.  However, the number of corrected variables is too few to cover all the flight envelop.

So, to enlarge the number of corrected variables, neural network is introduced to predict component maps. In \cite{yu2007neural}, a three layer BP neural network was trained to predict the compressor map using data provided by manufacturers. In \cite{ghorbanian2009artificial}, various neural networks were proposed to predict compressor performance using experimental data. Neural network based methods indeed increase the number of variables to be corrected. However, most neural networks take component characteristics modeling as a pure regression problem, ignoring the dynamics constrains during components working process.  In this case, these methods can only fit limited data from component characteristics test rig and serve as alternative methods when component maps cannot accessed. For example, in \cite{yu2007neural}, neural network is only trained by component performance data from manufacturers and in \cite{ghorbanian2009artificial}, only 42 experimental data of one compressor were used to train the network.

To extract more accurate component characteristics from big flight data and model gas turbine dynamics as a whole, we propose a thermodynamic based and flight data driven hybrid network for gas turbines modeling. Different from component maps based thermodynamic models, our network reconstructs the component characteristics in a data-driven way to take component degeneration and individual difference into consideration. Moreover, different from data driven methods, in the training phase, physical based equations and analytical mathematical description are used to ensure that the optimization converges to gas turbine's dynamics process.

In our network, thermodynamic method is used as a backbone to make sure all stations can be computed and the overall flow path calculation is physically explainable. Component characteristics including LPC, HPC, burner, HPT, LPT, bypass, mixer and nozzle are modeled using the inlet-exit reflection by neural network. The inputs of the network for certain component are the state parameters of this component's inlet while the outputs are the state parameters of the exit. So the volume of our network is much bigger than component maps based models and the component characteristics can be real-time corrected using big flight data in the network's training phase. 

Test case is conducted with 26970 flight data acquired from a two spool mixed turbofan. The results show that the accuracy of our hybrid model measured by max T6 relative error can reach 7\%, 5\% better than thermodynamic model with map fitting tool \cite{sethi2013map} used in PROOSIS \cite{alexiou2007advanced, bala2007proosis} and 8\% better than data driven method with similar volume.

The main contributions of our work lie in the following aspects: 
\begin{itemize}
	\item A novel gas turbine flow path analysis method from both thermodynamic and data driven perspectives is introduced.
	\item A flight data driven component characteristics modeling method considering component degeneration and individual difference is proposed.
	\item A new loss function considering the physical process of turbines and a two-phase training process are proposed to ensure the optimization converges to turbine's dynamics process.
	\item Experiment is conducted on a huge number of flight data and the accuracy of our model is higher than both thermodynamic based and data driven gas turbine models.
\end{itemize}

The remaining sections are organized as follows. First, some related works about the thermodynamic model and neural network are briefly introduced. Next, the structure of our network is presented in a bottom-up view. A two phase training process is proposed to guarantee the optimization converges toward gas turbines dynamics. At last, experiments are conducted under flight data and comparisons are made between our network and other methods including map fitting based thermodynamic model and data driven neural network.

\section{Related Works}
\label{sec:related works}
In this section, a brief introduction is given about the thermodynamic model and neural network method.
\subsection{Thermodynamic Model}
\label{sec:thermo}
\begin{figure}
	\centerline{\includegraphics[width=\columnwidth]{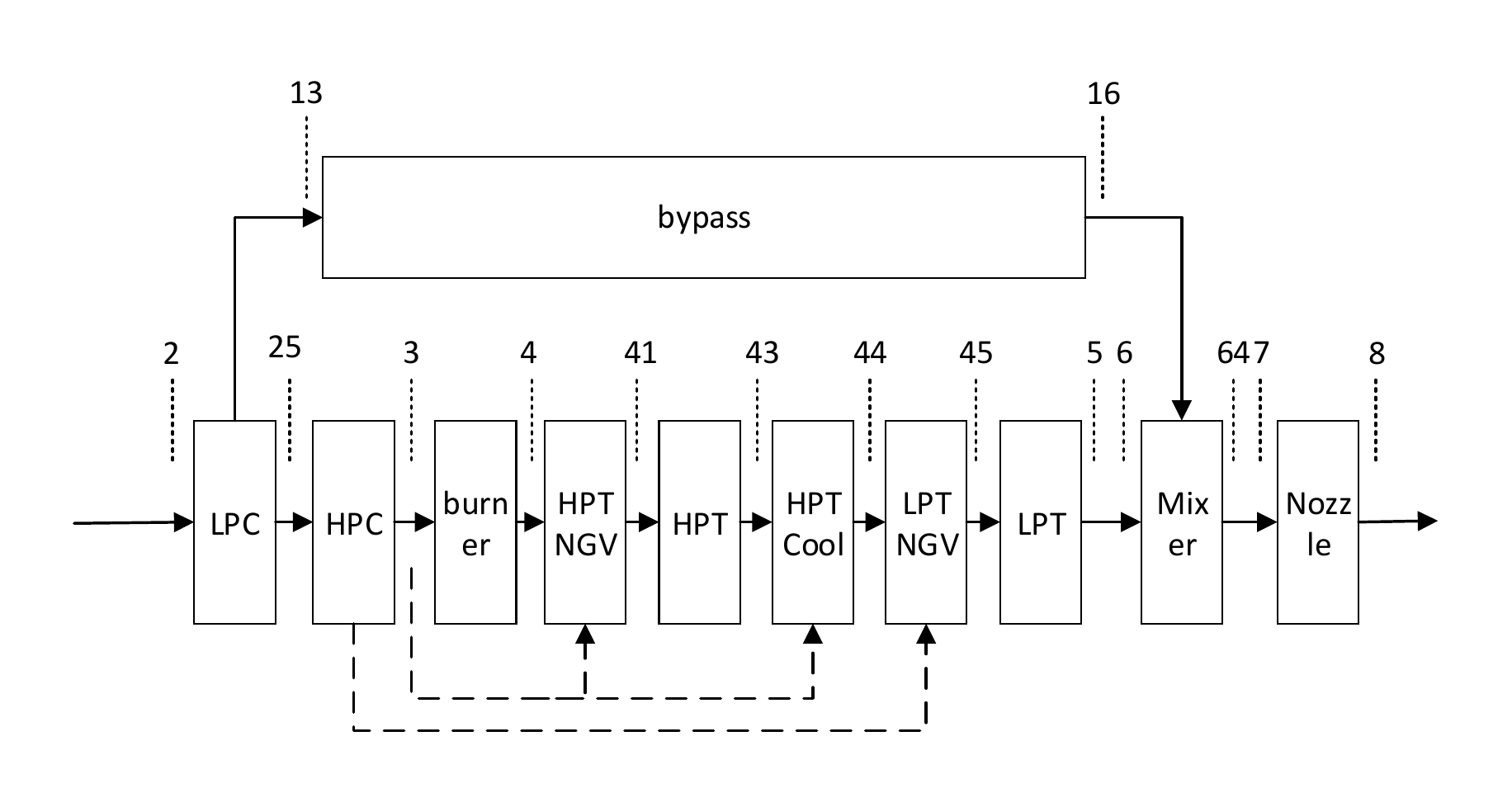}}
	\caption{The structure and station number of thermodynamic models}
	\label{fig:sttn}
\end{figure}

Thermodynamic model \cite{sellers1975dyngen,walsh2004gas} is a physical based method where the station state parameters are calculated from the inlet of the engine to the exit (Fig. \ref{fig:sttn}).  Main components such as compressor, burner and turbine are modeled by their characteristic maps acquired from component rig tests, simulation tests or corrected from standard maps. In the design point calculation, the  configuration such as the areas of some stations and the scale factors of component maps can be computed using the design point parameters. In the off-design point calculation, engine inlet parameters such as total temperature and pressure, the ambient pressure and some necessary engine control variables, such as N2, A8 and/or WF, are input into the model. Other parameters, such as N1, T4, beta values of component maps can be solved through thermodynamic balance equations' iterations. Once all variables are determined, the state parameters of each stations can be inferred by interpolating the component maps and thermodynamic calculation.

In thermodynamic model, parameters of all stations can be calculated and the performance of the engine can be easily analyzed. So it is widely used in the development cycle. As to the on-wing performance evaluation of gas turbine, the accuracy reduces with component characteristics drift because of various factors such as individual differences, degeneration and inlet-engine matching. Although some correction methods such as characteristic maps fitting are applied, the number of corrected variables is too few for the corrected map to cover all the flight envelop. 

\subsection{Fully Connected Neural Network}
\label{sec:fcnn}
Fully connected neural network \cite{hecht1992theory} is a data driven method including an input layer, a few hidden layers and an output layer.  Given a set $\Omega=\{(x_i, y_i)\}$ of input $x_i$ and its corresponding output $y_i$, neural network learns a reflection from $x_i$ to $y_i$ using affine transformations and nonlinear activation.

The dataset $\Omega$ is usually split into two subsets called training set $\Omega_{tr}$ and testing set $\Omega_{tt}$. In the training phase, a loss function $l(\hat{y_i}, y_i)$ is defined to measure the difference between the network output $\hat{y_i}$ and the real $y_i$ in the training set $\Omega_{tr}$. And the total loss in the training set is then defined:
\begin{equation}
loss=\sum_{(x_i, y_i)\in\Omega_{tr}}l(\hat{y_i}, y_i)=\sum_{(x_i, y_i)\in\Omega_{tr}}l(net(x_i), y_i).
\label{eq:trainingloss}
\end{equation}

Using gradient descent method, the weights and bias of the network are optimized toward the loss reduction direction. And after several iterations, the error can be tolerable if the learning rate is set reasonable. 

In the testing phase, the difference between $y_i$ in the testing set $\Omega_{tt}$ and its evaluation $\hat{y_i}$ by the trained network is used to measure the performance/accuracy of the network.

Compared with thermodynamic model, neural network has the following advantages.
\begin{itemize}
	\item Efficient. No iteration is needed during testing phase, so it is efficient in on-wing real-time calculations.
	\item Customized. If trained properly, the network can extract the individual characteristics and degeneration trend from the big flight data.
\end{itemize}

However, the disadvantages of the neural network are also obvious.
\begin{itemize}
	\item Hard to explain. Only affine transformations and activation can be accessed in the structure of the neural network, so the physical process in calculating the output is unclear.
	\item Hard to train. The training process pays no attention to other parameters not in flight data. So, the network is easy to run into over-fitting in the training set and gets poor performance in the testing set.
	\item Hard to analyze. Key station parameters not recorded in the flight data but essential to performance analysis cannot be evaluated by data driven methods.
\end{itemize}

From \ref{sec:thermo} and \ref{sec:fcnn}, we can conclude that thermodynamic model can calculate all parameters of key stations but can hardly model the engine's degeneration trend and individual characteristics with on-wing flight data. Neural network methods can construct customized model using big flight data but are hard to train with few constrains. So, a hybrid model taking thermodynamic models and physical based constrains as backbone and using neural network to model component characteristics can take their advantages while get rid of their disadvantages. This is why we propose our hybrid network.

\section{Proposed hybrid network}
\label{sec:turbinet}
\begin{figure}[htb]
	\centerline{\includegraphics[width=\columnwidth]{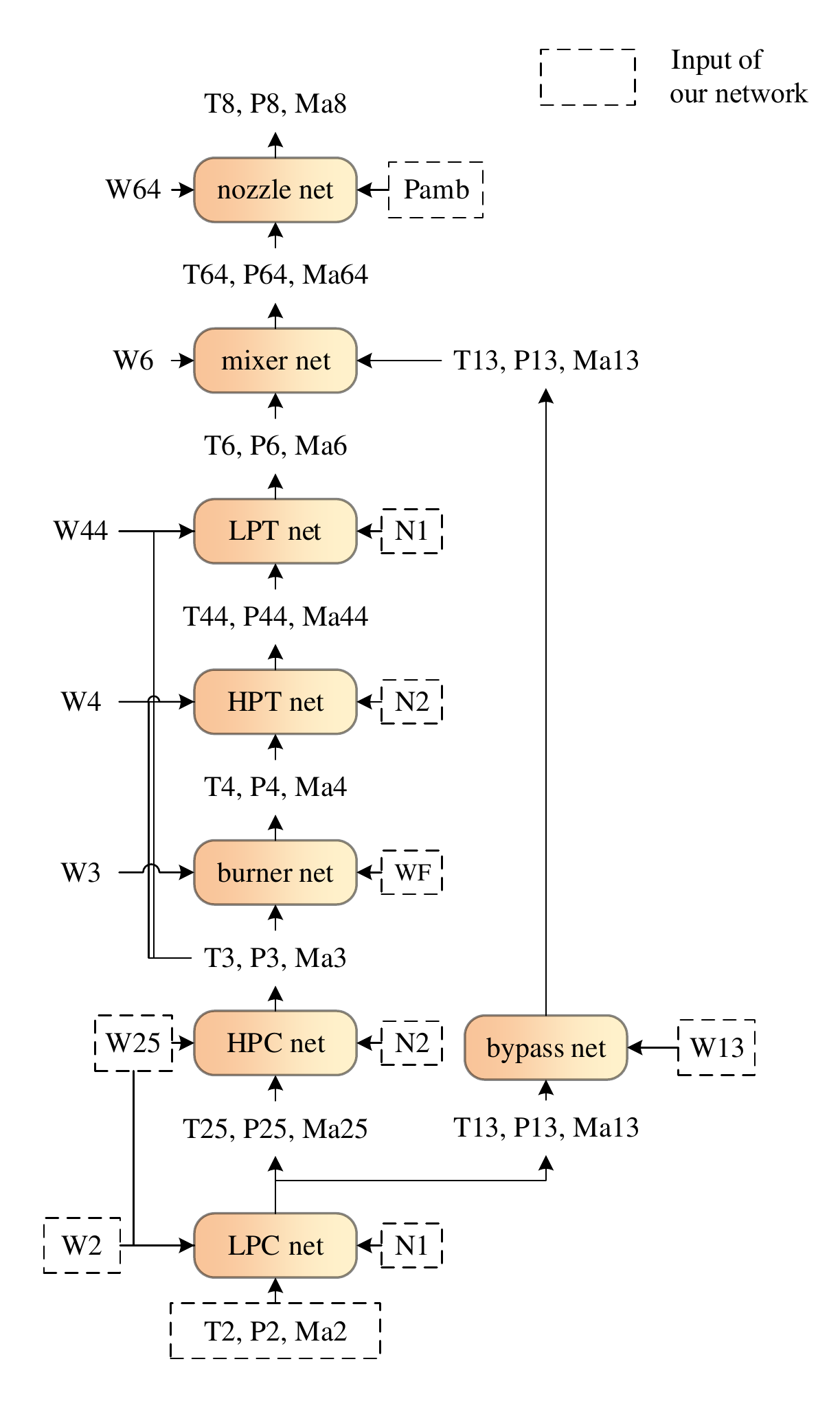}}
	\caption{The structure of our network}
	\label{fig:TurbiNet}
\end{figure}

In this section, we propose a hybrid network for gas turbine on wing modeling. The structure  is illustrated in Fig. \ref{fig:TurbiNet} and the station numbers in Fig. \ref{fig:TurbiNet} are the same as Fig. \ref{fig:sttn}. The input parameters are N1, N2, Wf, Pamb, T2, P2, Ma2, W2 and W25, the outputs are station state parameters such as total temperature and pressure. The component nets (described in Sec. \ref{sec:component net}) take the inlet parameters as inputs and output the exit parameters.  At last, the component nets are cascaded as the turbine's gas flow process in Fig. \ref{fig:sttn}.

Similar to thermodynamic model, our network is a component based model that calculates the state parameters station by station from the inlet of the engine. The main difference lies in that our network uses neural network to model every component other than component maps based thermodynamic calculation.

\subsection{Components Modeling}
\label{sec:component net}
\begin{figure}[htb]
	\centerline{\includegraphics[width=0.6\columnwidth]{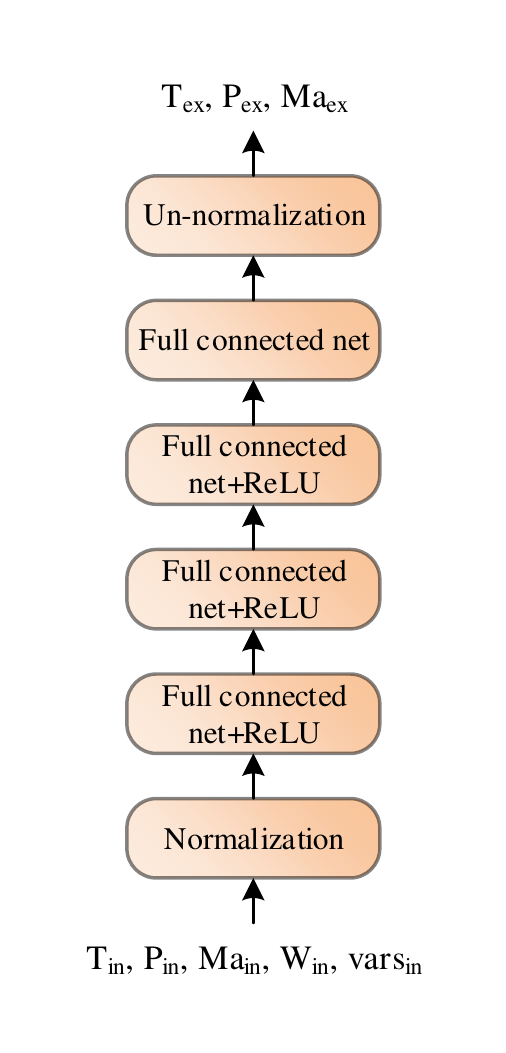}}
	\caption{The structure of base component net}
	\label{fig:basenet}
\end{figure}

The base component net is a four layers fully connected network (Fig. \ref{fig:basenet}) with ReLU activation funcion and all component nets inherit from it. The inputs of certain component net are the state parameters of component inlet plus other variables necessary to calculate the exit parameters. The outputs are state parameters of the exit station. To one station, under the assumption of one dimension steady flow, only four state parameters are needed to determine the station's other parameters. So for the convenience of calculation, total temperature T, total pressure P, mach number Ma and mass flow W are selected as the inputs of the inlet station. Because the exit mass flow can be inferred from the inlet mass flow and bleeding ratio, the outputs of the exit station are the three state parameters: total temperature T, total pressure P, mach number Ma. 

\begin{equation}
T_{ex}, P_{ex}, Ma_{ex}=net(T_{in}, P_{in}, Ma_{in}, W_{in}, vars),
\label{eq:basenet}
\end{equation}
where the function $net(\cdot)$ is certain component net and $vars$ are other variables needed to calculate the exit parameters. The following analyzes the required inputs of every component to compute the exit parameters assuming the component characteristics have been learned by network.

\subsubsection{HPC Net}
For HPC with one single inlet and one exit, another control variable, usually spool speed for its accessibility, is needed to calculate the exit parameters.

\subsubsection{LPC Net}
For LPC of turbofan with one inlet and two exits, there is a little difference from HPC. In HPC, the exit state is determined given the inlet parameters and spool speed. However, to LPC of turbofan, the exit state of LPC also depends on the bypass ratio. So mass flow of HPC inlet is added to inputs.

\subsubsection{Burner Net}
Fuel flow or one exit state parameter is needed to model burner. In most cases, fuel flow is easier to access. So in our network, fuel flow is added to the inputs of burner net.

\subsubsection{HPT and LPT nets with Cooling System}
If cooling system exists in the HPT or LPT, the simple rotor component nets such as HPC net is not sufficient. Because the cooling air from HPC to HPT and LPT can form a two inlets and one exit structure. To model turbines with cooling system, the inputs should include state parameters and mass flow of cooling air.

\subsubsection{Bypass Net}
Bypass is just a simple flow duct. Only three inlet state parameters and mass flow are required to calculate the exit parameters.

\subsubsection{Mixer Net}
To mixer, station parameters and mass flow of two inlets are all needed to determine the exit parameters.

\subsubsection{Nozzle Net}
The exit parameters of nozzel do not only depend on the inlet ones but also the ambient pressure. So, the input to nozzle net should also include the ambient pressure.

\subsubsection{Inputs and outputs of our network}
The component nets modeled above are cascaded to form our network (Fig. \ref{fig:TurbiNet}). The required inputs of our network are T2, P2, Pamb, Ma2, N1, N2, mass flow of each component inlet and WF. The outputs are state parameters of key stations. However, the mass flow of each component can be inferred only if the initial two mass flow W2 and W25 are given:
\begin{equation}
\begin{split}
W3&=W25\times (1-LNGV_{cl})+WF,\\
W4&=W25\times (1-HPT_{cl}-LNGV_{cl}-HNGV_{cl})+WF,\\
W41&=W4+W25\times HNGV_{cl},\\
W43&=W41,\\
W44&=W4+W25\times (HNGV_{cl}+HPT_{cl}),\\
W45&=W44+W25\times LNGV_{cl},\\
W5&=W45,\\
W6&=W5,\\
W13&=W2-W25,\\
W16&=W13+c2b\times W25,\\
W64&=W2+WF,\\
W8&=W64,
\end{split}
\label{eq:W_infer}
\end{equation}
where $HPT_{cl}$ is the HPT cooling proportion of W25 from HPC,  $LNGV_{cl}$ is the LPT NGV cooling proportion of W25 from HPC, $HNGV_{cl}$ is the HPT NGV cooling proportion of W25 from HPC and $c2b$ is the leakage proportion of W25 from the core to bypass. So our network inputs can be reduced to T2, P2, Ma2, Pamb, N1, N2 W2, W25 and WF. In Tab. \ref{tab:component net}, we give a brief description about the inputs and outputs of every component net.

\begin{table}
	\centering
	\caption{Inputs and outputs of every component net}
	\label{tab:component net}
	\setlength{\tabcolsep}{3pt}
	\begin{tabular}{p{50pt}|p{80pt}|p{90pt}}
		\hline
		component& inputs& outputs\\
		\hline
		LPC& T2 P2 Ma2 \par W2 W25 N1& T25 P25 Ma25 \par T13 P13 Ma13\\
		\hline
		HPC& T25 P25 Ma25 W25 N2& T3 P3 Ma3\\
		\hline
		burner& T3 P3 Ma3 W3 WF & T4 P4 Ma4\\
		\hline
		HPT& T4 P4 Ma4  W4\par T3 P3 Ma3 N2 & T44 P44 Ma44\\
		\hline
		LPT& T44 P44 Ma44 W44\par T3 P3 Ma3 N1&T6 P6 Ma6\\
		\hline
		bypass& T13 P13 Ma13 W13& T16 P16 Ma16\\
		\hline
		mixer& T6 P6 Ma6 W6 \par  T16 P16 Ma16 W16& T64 P64 Ma64\\
		\hline
		nozzle& T64 P64 Ma64 W64 Pamb& T8 P8 Ma8\\
		\hline
		 the whole network& T2 P2 Ma2 Pamb \par N1 N2 W2 W25 WF& parameters above\\
		\hline
		\multicolumn{3}{p{251pt}}{ The station number is defined in Fig. \ref{fig:sttn}.}\\
	\end{tabular}
\end{table}

\begin{figure}[htb]
	\centerline{\includegraphics[width=\columnwidth]{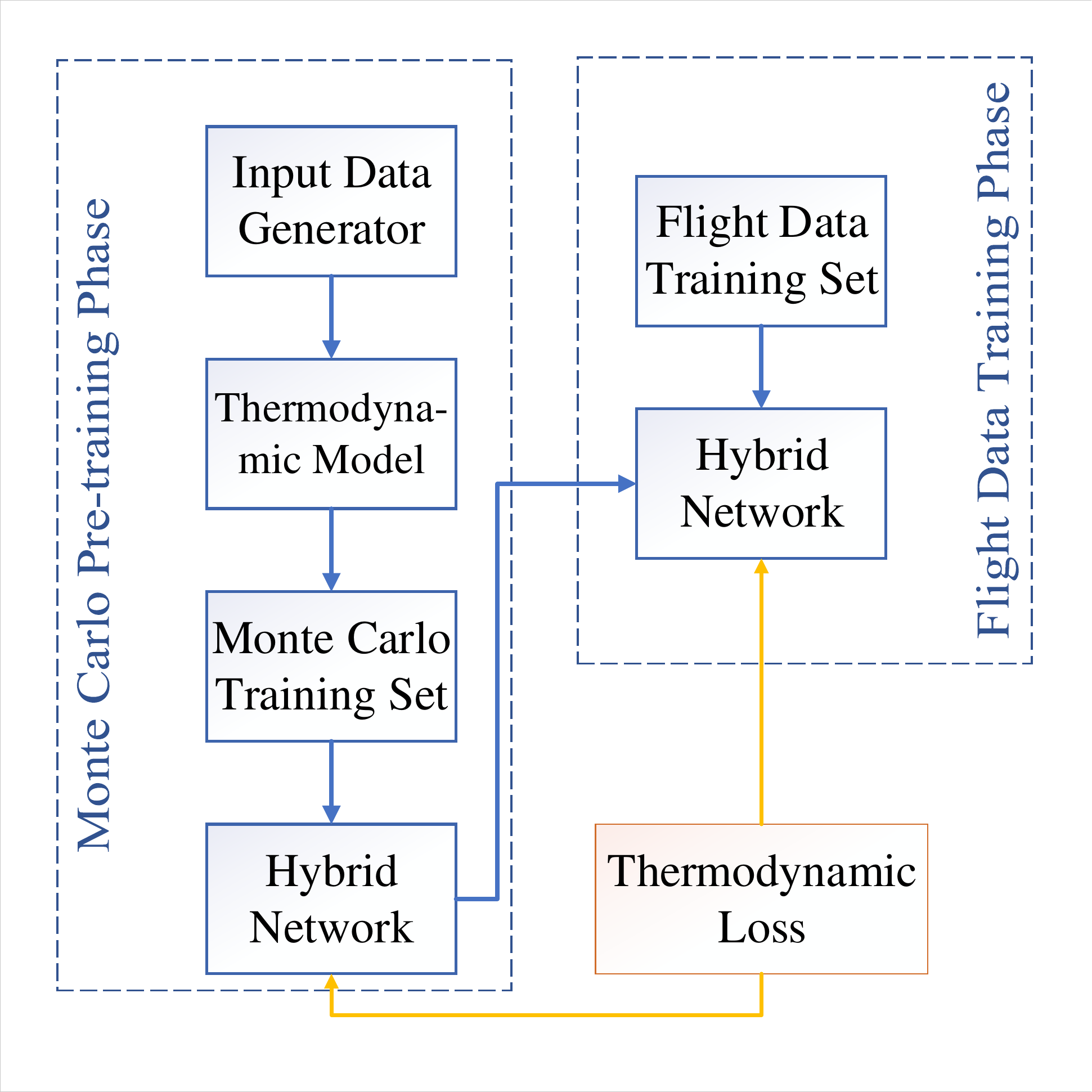}}
	\caption{The two phase training process of our network}
	\label{fig:flowchart}
\end{figure}

Once our network is constructed, the next issue is how to train it to model component characteristics. In the next section,  a two phase training method is proposed (Fig. \ref{fig:flowchart}).

In the Monte Carlo pre-training phase, Monte Carlo training set is generated with inputs from Monte Carlo simulation and outputs from thermodynamic model. our network is trained using Monte Carlo training set to model approximate component characteristics. In the flight data training phase, our network is trained with flight data to correct the component characteristics. Thermodynamic constrains are intervened during optimization to guarantee  the training phase converges to the dynamics process.

\subsection{Monte Carlo Pre-training Phase}
\label{sec:mc}
Most of data driven gas turbine modeling methods define the loss function as the error between the model's outputs and the real values collected from sensors. Subjected to the limited number of gas turbine sensors, the loss function is only the errors of a few parameters such as T6. In this condition, the optimization will try its best to regress these parameters in the loss function. But meanwhile, the measurement and random errors are also regressed.  Moreover, the component nets error will transmitted to the final outputs under measurement error loss. 

\begin{figure}[htb]
	\centerline{\includegraphics[width=\columnwidth]{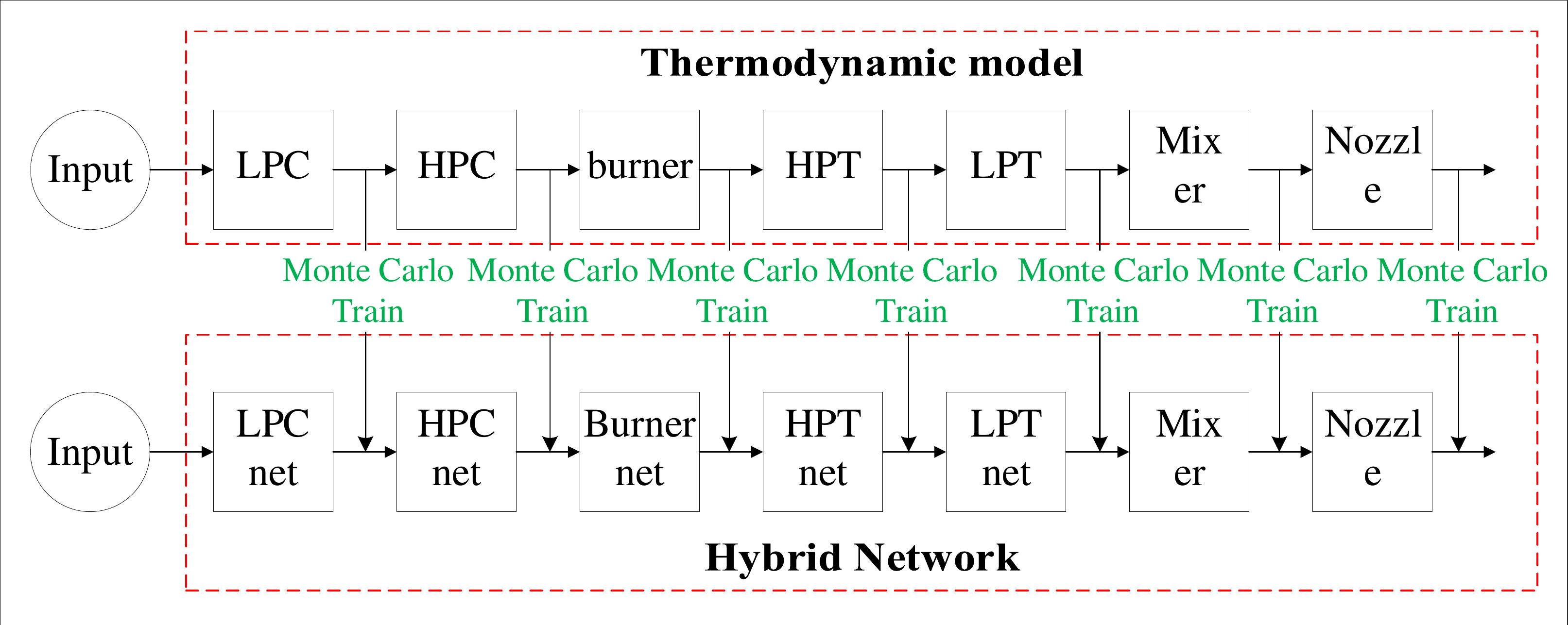}}
	\caption{ The schematic diagram of the Monte Carlo pre-training phase}
	\label{fig:MC}
\end{figure}

To avoid this phenomena, a Monte Carlo pre-training phase and thermodynamic loss are proposed. Dataset including all station parameters is first generated using Monte Carlo simulation and thermodynamic calculation. Then, our network is trained to output all these parameters in this dataset with the constrain of thermodynamic loss. When the error accumulated, the thermodynamic loss will increase and the training process will adjust the weights to decrease overall error. The schematic diagram of the Monte Carlo pre-training phase is illustrated in Fig. \ref{fig:flowchart} and Fig. \ref{fig:MC} and the bypass part is omitted in Fig. \ref{fig:MC}  for simpleness.

First, a number of inputs required by thermodynamic model, usually T2, P2, Pamb and N2 are generated randomly in a reasonable scope in a Monte Carlo simulated test way.
\begin{equation}
T2_i, P2_i, Pamb_i, N2_i=random(), i=1,2,\cdots,N
\label{eq:inputGen}
\end{equation}
where the subscript $i$ denotes the $i$th samples.

Then, these variables are input into thermodynamic model to calculate all variables and stations parameters required by our network's inputs and outputs.
\begin{equation}
\begin{split}
Ma2_i, N1_i, W2_i, W25_i, WF_i, sttns_i\\
=thermo(T2_i, P2_i, Pamb_i, N2_i),
\end{split}
\label{eq:thermoCal}
\end{equation}
where $sttns_i$ are the stations parameters required by our network's outputs.

When the inputs and outputs of our network are all computed, the dataset $\Omega_{mc}=\{(x_i, y_i), i=1,2,\cdots,N\}$ can be generated. In  $\Omega_{mc}$, $x_i$ is the input vector of the $i$th samples for our network and $y_i$ is the output vector of our network.

To decrease the influence of error transmission from each component net, we define thermodynamic loss to measure the distance between the net's outputs and the turbine's dynamic. Once the error accumulated, the network's process will deviated from the turbine's dynamics, so the  thermodynamic loss calculated by the network outputs will increase. Under the constrain of thermodynamic loss, the overall error of our hybrid network will not accumulated. 

Next is the pre-training process. The loss function is optimized to decrease using gradient descend method. In traditional network based modeling methods, loss function is usually defined as the error between the real values and their evaluations by network. However, to ensure the optimization converges toward the real dynamics process of gas turbines and to decrease the influence of error transmission from each component net, new loss measuring the error between our network's calculation process and real thermodynamic process is introduced. Once the error accumulated, the network's calculation process will deviated from the turbine's dynamics and the  thermodynamic loss calculated by the network outputs will increase. So under the constrain of thermodynamic loss, the overall error of our hybrid network will not accumulated. So, in our network, the loss function is comprised of two parts: one is stations parameters errors loss and the other one is the thermodynamic error loss.

The first loss is defined as mean square error between the outputs of our network and the corresponding stations parameters calculated by thermodynamic model:
\begin{equation}
loss_1=\frac{1}{N}\sum_{(x_i, y_i)\in\Omega_{mc}}\frac{(Net(x_i)-y_i)^2}{y_i^2},
\label{eq:TurbiNetloss}
\end{equation}
where $N$ is the number of samples in $\Omega_{mc}$.

The second loss is the  thermodynamic error loss measuring our network's deviation from gas turbine's dynamics. The loss is comprised of two parts: the mass flow loss and the power loss. The  mass flow loss is the error between mass flow of all stations computed by our network and mass flow inferred from inputs W2 and W25.
\begin{equation}
loss_2=\frac{1}{N}\frac{1}{M}\sum_{\Omega_{mc}}\sum_{i\in sttns}\frac{(W_i-Q(T_i,P_i,Ma_i,A_i))^2}{W_i^2},
\label{eq:Weqs}
\end{equation}
where $M$ is the number of stations in our network, $sttns$ denotes the stations number set accessed from our network, $(T_i,P_i,Ma_i)$ are the total temperature, total pressure and Mach number in our network's $i$th output station, $A_i$ is the $i$th station area accessed in design point calculation of thermodynamic model, $W_i$ can be inferred from W2 and W25 using Eq. \ref{eq:W_infer} and $Q(\cdot)$ is the mass flow function:
\begin{equation}
Q(T_i,P_i,Ma_i,A_i)=K\frac{P_iA_i}{\sqrt{T_i}}q(Ma_i)
\label{eq:Qfunction}
\end{equation}

The power loss is the error between compressor power demanded and turbine power provided.
\begin{equation}
\begin{split}
loss_3=&\frac{1}{N}\sum_{\Omega_{mc}}\frac{|\eta_l\Delta H_{LPT}-\Delta H_{LPC}|^2}{\Delta H_{LPT}^2}\\
+&\frac{|\eta_h\Delta H_{HPT}-\Delta H_{HPC}-P_{ext}|^2}{\Delta H_{HPT}^2},
\end{split}
\label{eq:powereqs}
\end{equation}
where $\Delta H_{LPT}$ is the enthalpy drop of LPT, $\Delta H_{HPT}$ the enthalpy drop of HPT, $\Delta H_{LPC}$ the enthalpy rise of LPC, $\Delta H_{HPC}$ the enthalpy rise of HPC, $\eta_h$ the mechanical efficiency of HPT, $\eta_l$ the mechanical efficiency of LPT and $P_{ext}$ the power extraction from HP spool:
\begin{equation}
\begin{split}
\Delta H_{LPT}&=H44+H_{LNGVcl}-H6,\\
\Delta H_{HPT}&=H4+H_{HNGVcl}-(H44-H_{HPTcl}),\\
\Delta H_{HPC}&=H3+H_{LNGVcl}-H25,\\
\Delta H_{LPC}&=H13+H25-H2,\\
\end{split}
\label{eq:thermoCal}
\end{equation}
where all the enthalpy above can be calculated using W2, W25 plus stations parameters from our network's outputs.

Then, our network is trained using gradient descend method toward reducing the sum of these three losses. If trained properly, our network can learn an approximate component characteristics and perform as good as thermodynamic models in calculating the stations parameters.

\subsection{Flight Data	Training Phase}
\label{sec:fdtrain}
Because Monte Carlo dataset is generated by thermodynamic model, the accuracy of our network can not be higher than thermodynamic model after Monte Carlo pre-training phase. To improve our network's accuracy, the flight data training phase is proposed to correct component characteristics by adjusting our network's weights using flight data.

First, quasi steady state is defined where the amplitude of N1 is less than $1\%$ during 3 seconds without consideration of the fluctuation of other variables such as Mach number, altitude and inlet temperature. This is a very relaxed quasi steady state determinant conditions and poses great challenges to performance calculation. 

Next, quasi steady state points are extracted from certain engine during a period to form dataset $\Omega_{fd}=\{(\overline{x_i}, y_i), i=1,2,\cdots,N\}$. W2 and W25 are usually not measured, so the inputs to our network are not sufficient. To remedy this, we use thermodynamic model to calculate this two variables (this two variables can also be calculated through our network's iteration phase proposed in \ref{sec:iter}):
\begin{equation}
W2_i, W25_i=thermo(\overline{x_i}).
\label{eq:w2w25}
\end{equation}
Then, we added this two variables into $\overline{x_i}$ to form the final flight dataset $\Omega_{fd}=\{(x_i, y_i), i=1,2,\cdots,N\}$.

The loss function of the flight data training phase is also comprised of two part as introduced in \ref{sec:mc}. The main difference lies in the parameters errors loss. Because there are only a few stations parameters in flight data, so not all the stations parameters output by our network will contribute to the loss. The parameters errors loss is defined as:
\begin{equation}
loss_1=\frac{1}{N}\sum_{(x_i, y_i)\in\Omega_{fd}}\frac{(S(Net(x_i))-y_i)^2}{y_i^2},
\label{eq:TurbiNetloss}
\end{equation}
where $S(Net(x_i))$ extract network's corresponding outputs to $y_i$ in flight data to compute the loss.

The thermodynamic error loss is the same as Monte Carlo pre-training phase defined in Eq. \ref{eq:Weqs} and \ref{eq:powereqs}. Then, our network is trained using gradient descend method to reduce the sum of these three losses in the flight dataset.  After flight data training phase, component characteristics are corrected using the on-wing flight data. So, the modeling accuracy will be higher than our network only trained by Monte Carlo dataset as well as thermodynamic models.

\subsection{Iterating Phase}
\label{sec:iter}
Usually, among the inputs of our network, T2, P2, Pamb, Ma2, N1, N2 and WF can be acquired from flight data. However, most flight data will not record the mass flow of LPC and HPC inlet: W2 and W25. In \ref{sec:fdtrain}, we take a compromised way by using thermodynamic model to calculate them. In this section, an optimization method is proposed to get rid of thermodynamic model during testing phase.

In the iterating phase, the weights and bias of our network are fixed and no longer optimized. W2 and W25 are the only two variables to be optimized. The optimization goal is comprised of two part: the mass flow error and  the power error.

The first optimization goal is to minimize the mass flow error described in Eq. \ref{eq:Weqs}. The loss can be modified into:
\begin{equation}
\begin{split}
loss_1&=f_1(W2,W25,Net(vars))\\
&=\frac{1}{N}\frac{1}{M}\sum_{(x_i, y_i)\in\Omega_{fd}}\sum_{i\in sttns}\frac{(W_i-Q(T_i,P_i,Ma_i,A_i))^2}{W_i^2}
\end{split}
\end{equation}
where $Net(\cdot)$ denotes our network and $vars$ denotes the inputs of our network.

The second optimization goal is to minimize the power error described in Eq. \ref{eq:powereqs}. The loss can also be modified into:
\begin{equation}
\begin{split}
loss_2&=f_2(W2,W25,Net(vars))\\
&=\frac{1}{N}\sum_{(x_i, y_i)\in\Omega_{fd}}\frac{|\eta_l\Delta H_{LPT}-\Delta H_{LPC}|^2}{\Delta H_{LPT}^2}\\
&+\frac{|\eta_h\Delta H_{HPT}-\Delta H_{HPC}-P_{ext}|^2}{\Delta H_{HPT}^2}.
\end{split}
\end{equation}

Then we minimize the following goal regarding to W2 and W25:
\begin{equation}
\arg\min_{W2,W25}f_1(W2,W25,Net(vars))+f_2(W2,W25,Net(vars)).
\label{eq:min}
\end{equation}
Given an initial value for W2 and W25, we can get the optimal W2 and W25 through an iteration process.

\section{Test case}
In this section, the accuracy of our network is tested in flight data. Flight data are gathered from a two spool turbofan over a period. N1, N2, WF, T2, P2, Pamb are accessible for the inputs of our network. W2 and W25 can be either calculated by thermodynamic model introduced in \ref{sec:fdtrain} or iteration method in \ref{sec:iter}. Because neither Ma2 nor PS2 is measured in flight data, the Ma2 input of our network cannot accessed or computed. In this case, we simply omit the Ma2 input of LPC net under the assumption that the exit parameters of LPC can be determined by other inputs as long as the inlet area keeps unchanged. There is only one station parameter T6 recorded  in the flight data, so we use it to measure the accuracy of our network. Two datasets are made to the whole training and testing process: $\Omega_{mc}$ is generated using Monte Carlo method introduced in \ref{sec:mc} and $\Omega_{fd}$ is gathered as introduced in \ref{sec:fdtrain} and \ref{sec:iter}. 
\subsection{Monte Carlo Training Results}
our network is first trained by $\Omega_{mc}$ generated using Monte Carlo methods. The number of samples in $\Omega_{mc}$ is $15360$ and the ratio of the training set to the test set is $0.8$. The number of training epochs is $500$, the learning rate is $1e-3$ with $10\%$ decay every $100$ epochs and the batch size is set to $256$. The training loss is defined as the sum of Eq. \ref{eq:TurbiNetloss}, \ref{eq:Weqs} and \ref{eq:powereqs}. The accuracy is measured by the max relative error in all samples of training set.
\begin{equation}
max\_error=\max_{i\in \omega}\frac{|\hat{y_i}-y_i|}{y_i},
\label{eq:maxerror}
\end{equation}
where $y_i$ is certain station parameter of sample $i$ and $\omega$ is subscript set of dataset $\Omega_{mc}$

\begin{figure*}
	\centering
	\subfigure[Max error of Ma3]{\includegraphics[width=.45\columnwidth]{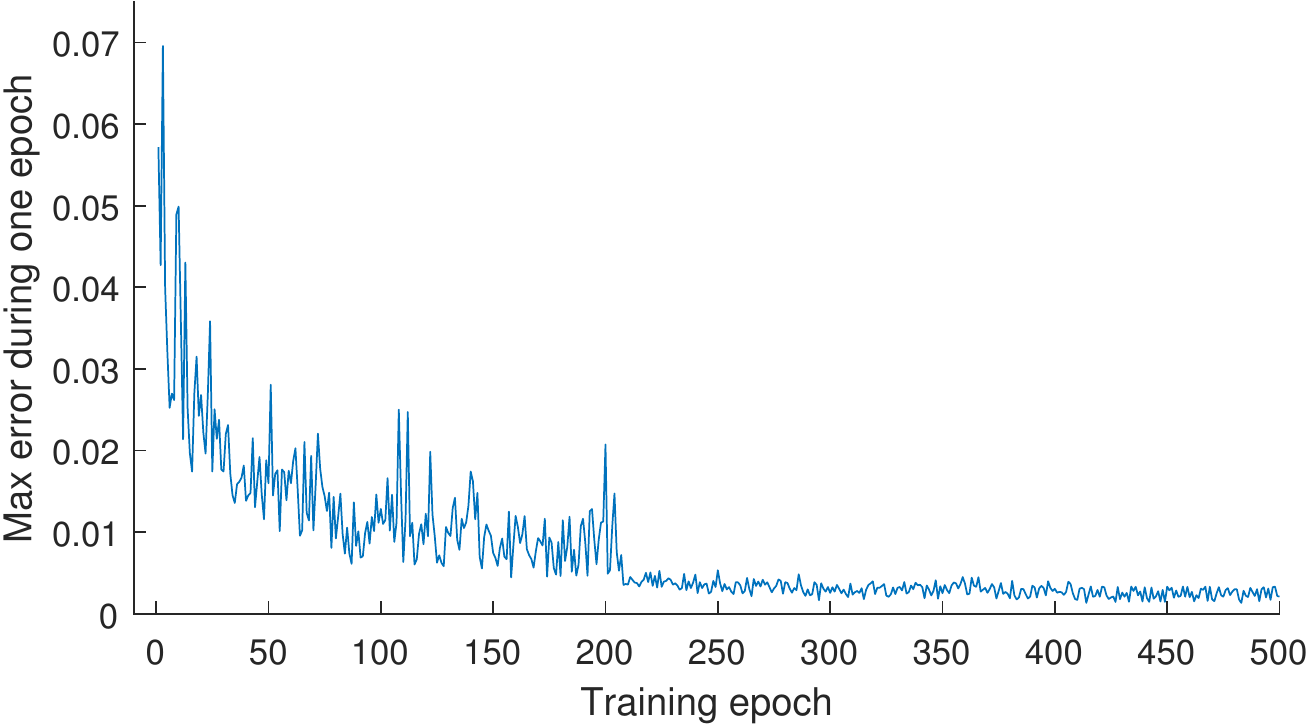}}
	\subfigure[Max error of P4]{\includegraphics[width=.45\columnwidth]{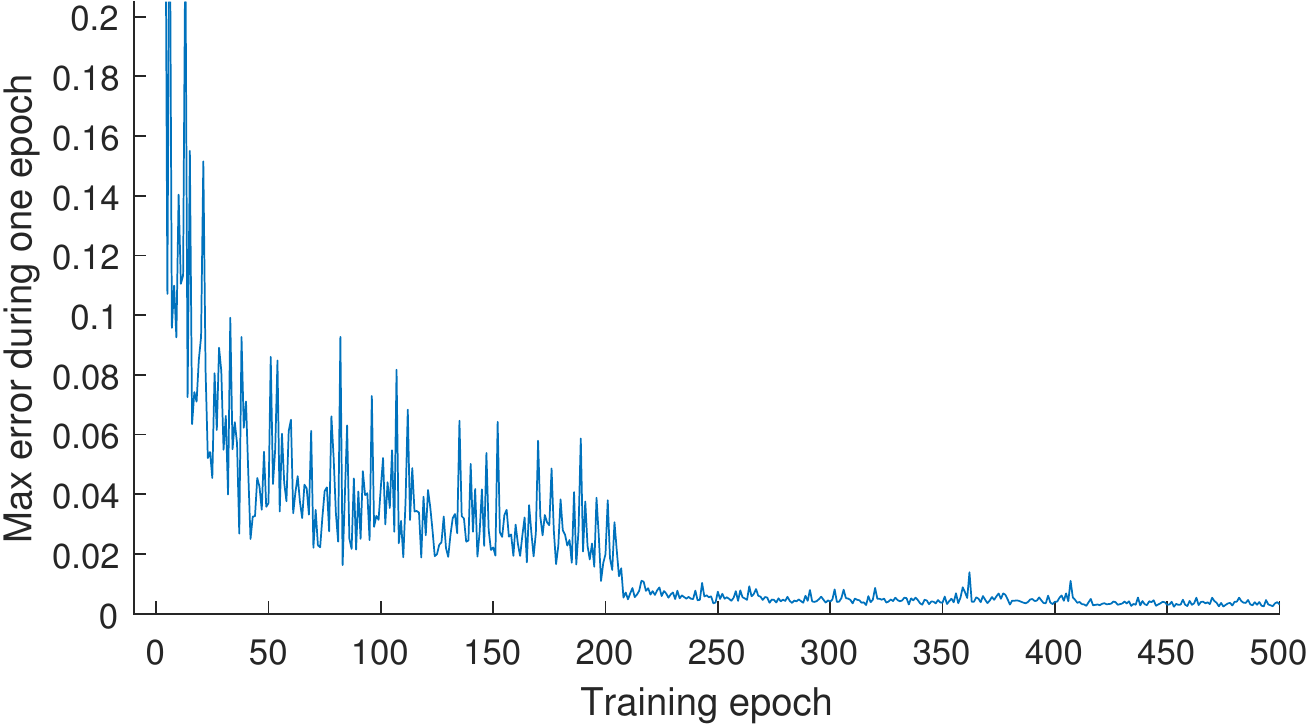}}
	\subfigure[Max error of T6]{\includegraphics[width=.45\columnwidth]{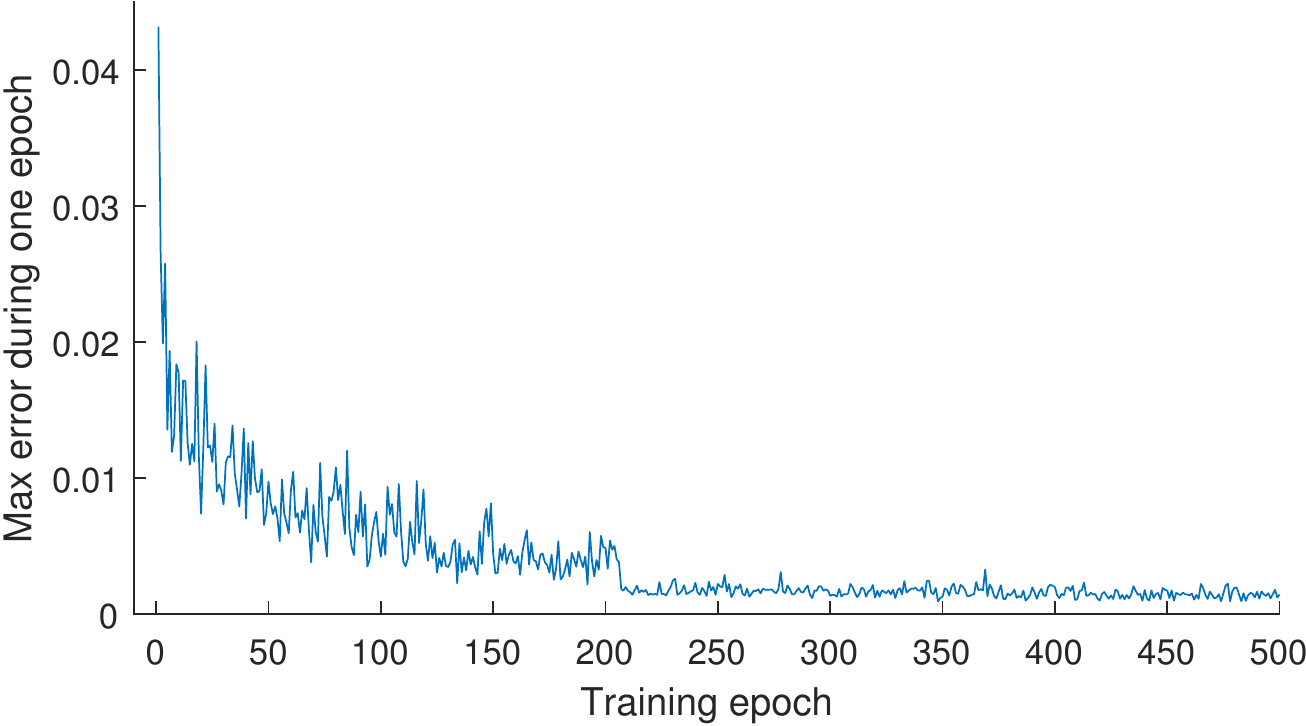}}
	\subfigure[Max error of T8]{\includegraphics[width=.45\columnwidth]{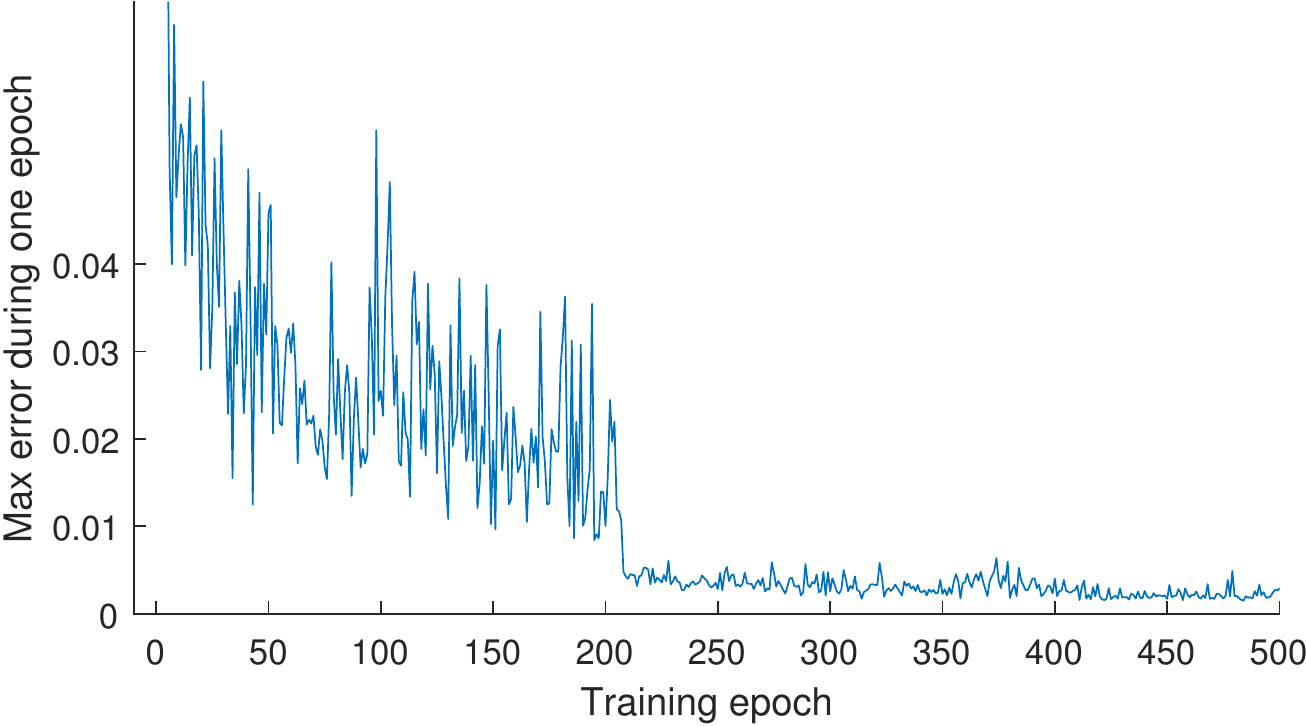}}
	\caption{The error trending in training process of some parameters}
	\label{fig:errorstrainprocess}
\end{figure*}

\begin{figure}[htb]
	\centerline{\includegraphics[width=0.9\columnwidth]{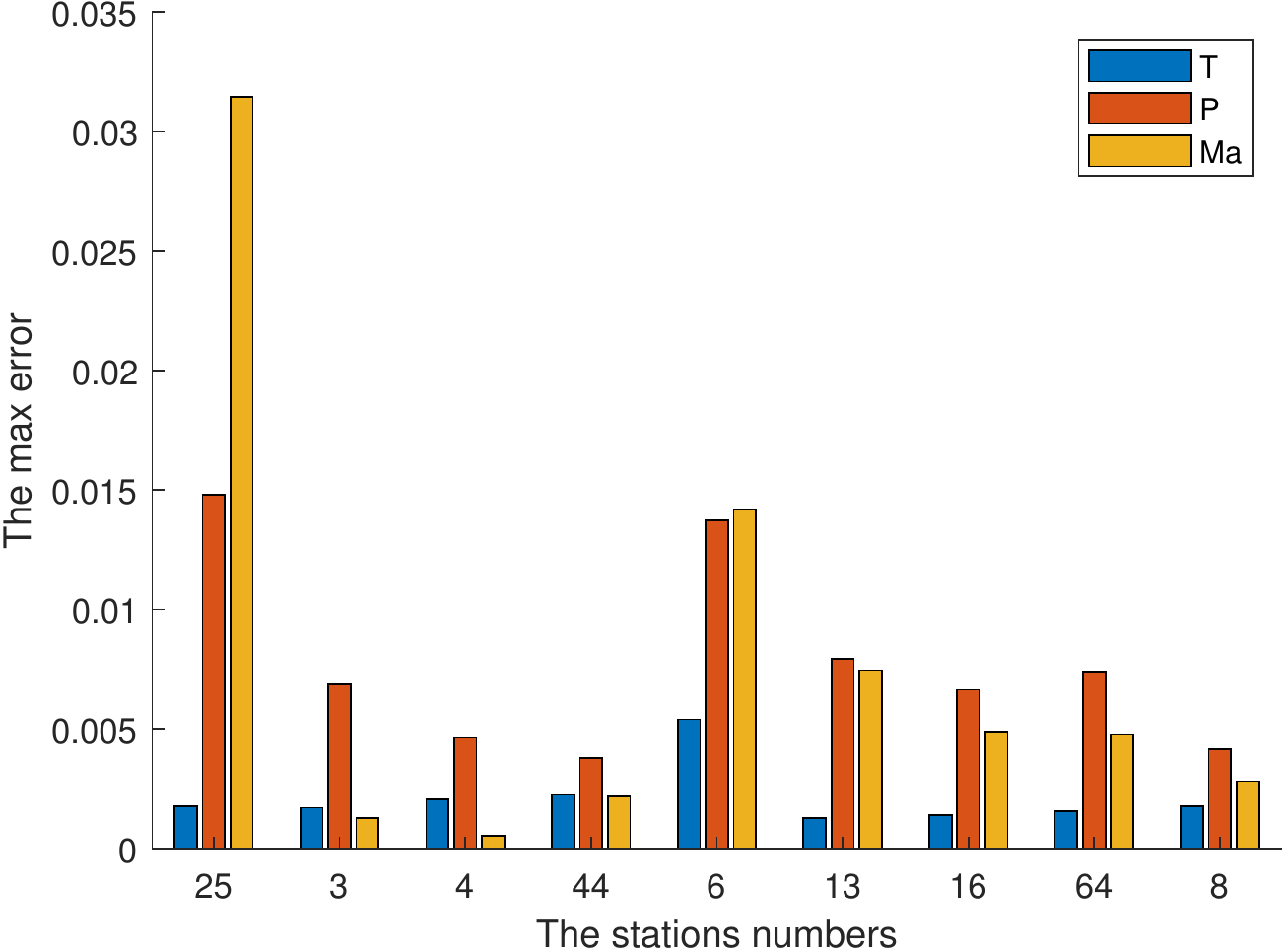}}
	\caption{The max relative errors after Monte Carlo training}
	\label{fig:MCerrors}
\end{figure}

Fig. \ref{fig:errorstrainprocess} shows the error trending of  Ma3, P4 and T6 in the training process. We can see that the training phase converges after about 200 epochs. The max relative errors of all stations parameters  after Monte Carlo training are described in Fig. \ref{fig:MCerrors}. The Ma error of station 25 (the LPC exit) is 3.2\% that is obviously larger than other stations. This is because the Ma2 input of LPC net is omitted. Except Ma25, the errors of other parameters are less than 1.5\%.

\subsection{Flight Data Training Results}
After trained by Monte Carlo dataset, our network is further trained with flight dataset as introduced in \ref{sec:fdtrain}. The flight dataset is split into two subsets: the training set $\Omega_{fd-tr}$ to train our network and testing set $\Omega_{fd-tt}$ to measure the accuracy and generalization of our network. There are $6,970$ samples in$\Omega_{fd-tt}$ and $20,000$ samples in $\Omega_{fd-tr}$.  Because there are only one station parameter T6 in flight data, so the parameters loss can be modified into:
\begin{equation}
loss=\frac{1}{N}\sum_{i\in \omega}\frac{(\hat{T6_i}-T6_i)^2}{T6^2},
\label{eq:T6loss}
\end{equation}
where $\hat{T6_i}$ is our network's evaluation  for T6 of the $i$th sample in $\Omega_{fd-tr}$,  $\omega$ is subscript set of training flight dataset $\Omega_{fd-tr}$ and $N$ is the number of samples in $\Omega_{fd-tr}$.

\begin{figure}[htb]
	\centerline{\includegraphics[width=0.9\columnwidth]{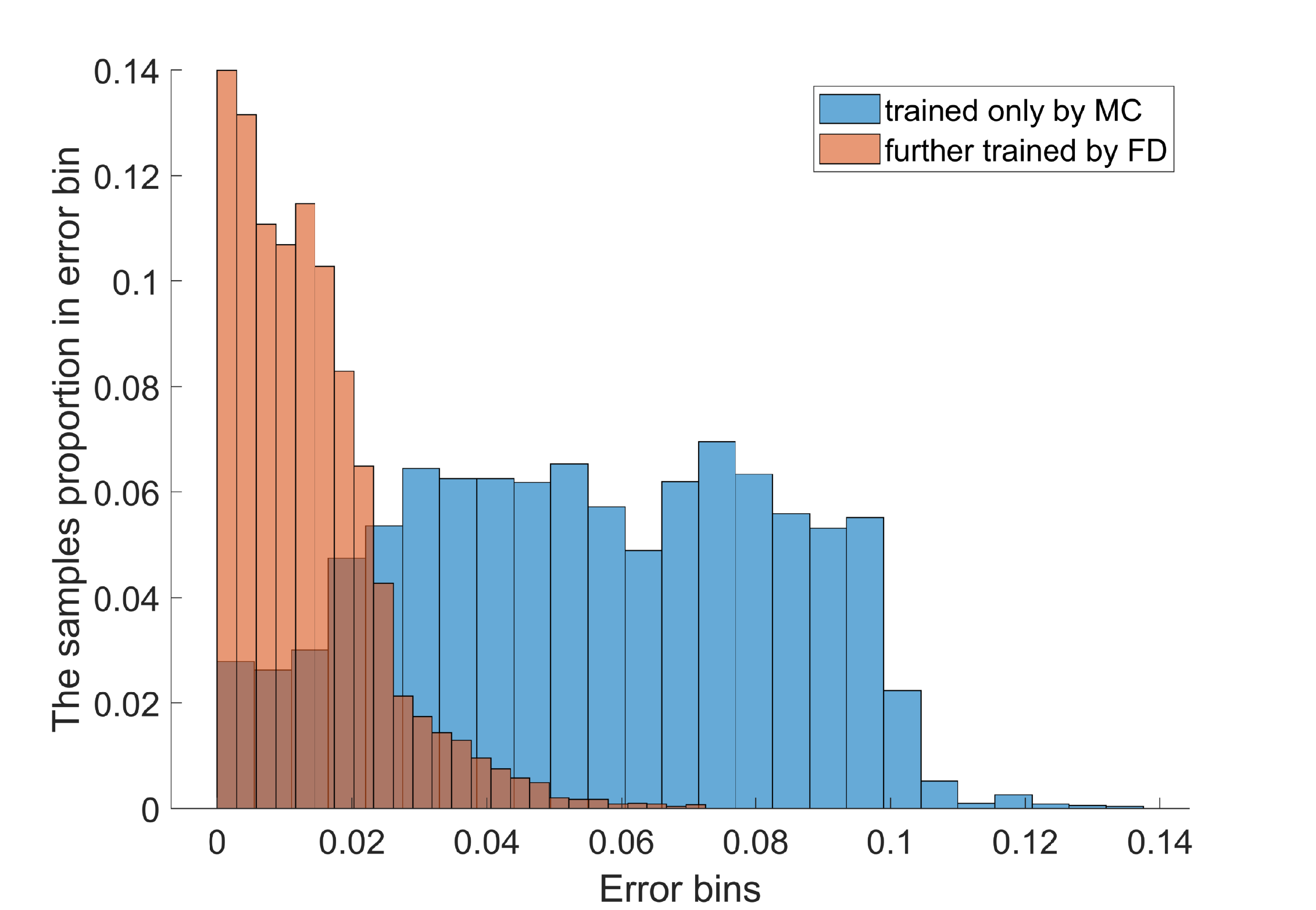}}
	\caption{T6 errors comparison between our network only trained by MC dataset and further trained by FD dataset}
	\label{fig:T6errorsMCFD}
\end{figure}

To compare the performance before and after flight data training, we evaluate T6 of flight testing dataset using our network only trained by Monte Carlo dataset first. Then, T6 is evaluated using our network further trained by flight training dataset. The T6 relative errors histogram of all samples in the testing set is shown in Fig. \ref{fig:T6errorsMCFD}.

We can see that before trained by flight data, the T6 relative errors' distribution is nearly uniformed around 6\% and the max relative error can reach 13.8\%. After further trained by flight data, the T6 relative errors mainly concentrated below 4\% and the max error reach 7.1\%. This conforms that after flight data training, our network has the ability of correcting component characteristics to get a more accurate evaluation. Another phenomena from Fig. \ref{fig:T6errorsMCFD} is that the error distribution forms a long tail shape. This long tail error distribution is much more suitable for gas turbine monitoring than uniform distribution. When T6 error distribution during a period deviated from this long tail distribution, attention or inspection should be taken. However, the criterion about whether the performance degenerates or latent failure occurs is beyond the scope of this paper.

\subsection{Iterating Results}
\begin{figure}[htb]
	\centerline{\includegraphics[width=.9\columnwidth]{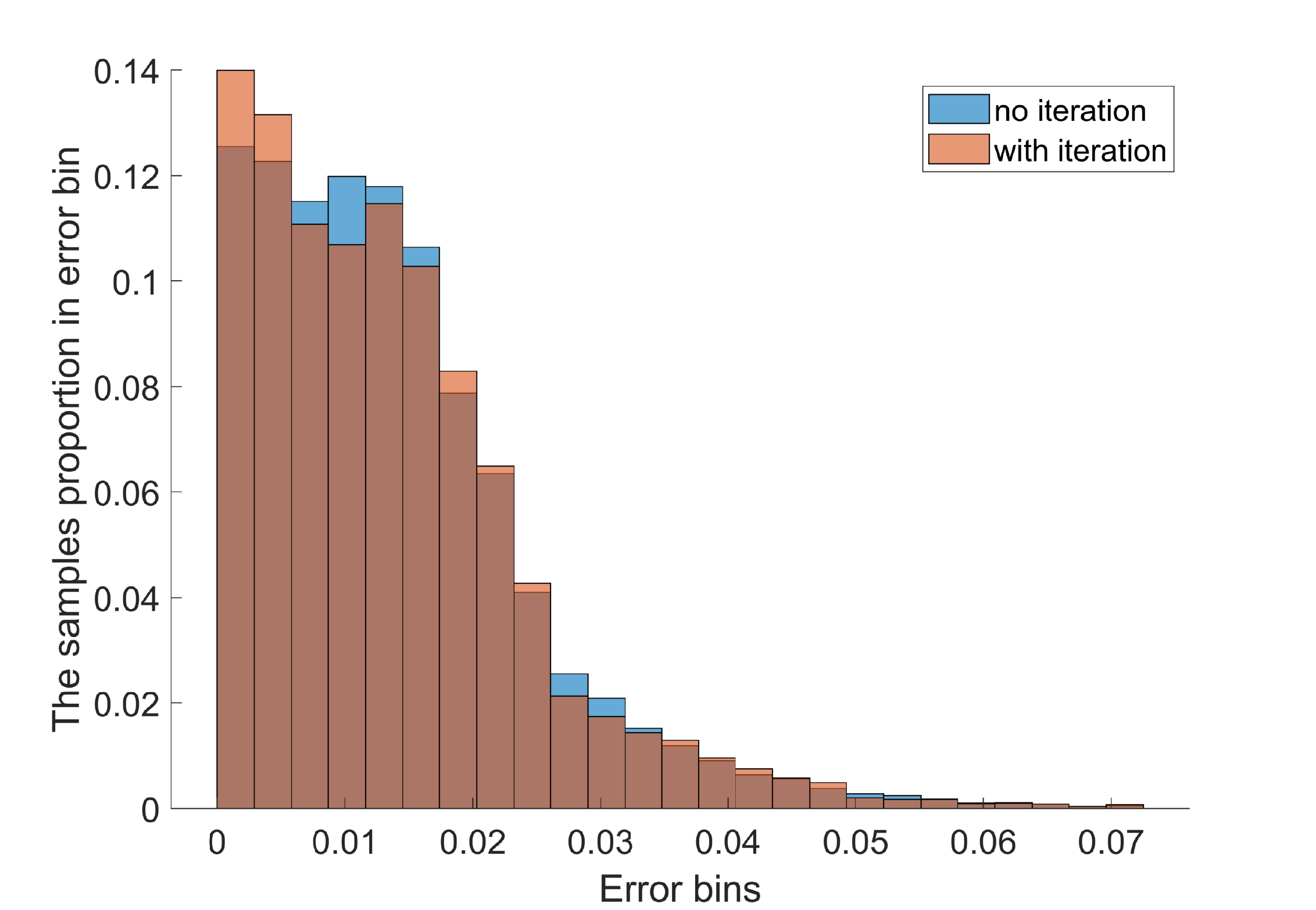}}
	\caption{T6 errors comparison between iteration and no iteration methods}
	\label{fig:T6errorsIter}
\end{figure}

Our network with iteration phase proposed in \ref{sec:iter} is also tested. The results are shown in Fig. \ref{fig:T6errorsIter}. The error distributions of iteration method and no iteration method are nearly the same. This verifies the effectiveness of iteration method introduced in \ref{sec:iter}. So our network can be independent from thermodynamic models where more iterative steps and calculation times are required. Then, our network will be more efficient and has prospects of being embedded in airborne equipment to realize real-time calculations.

\subsection{Comparisons with Thermodynamic Model}
\begin{figure}[htb]
	\centerline{\includegraphics[width=.9\columnwidth]{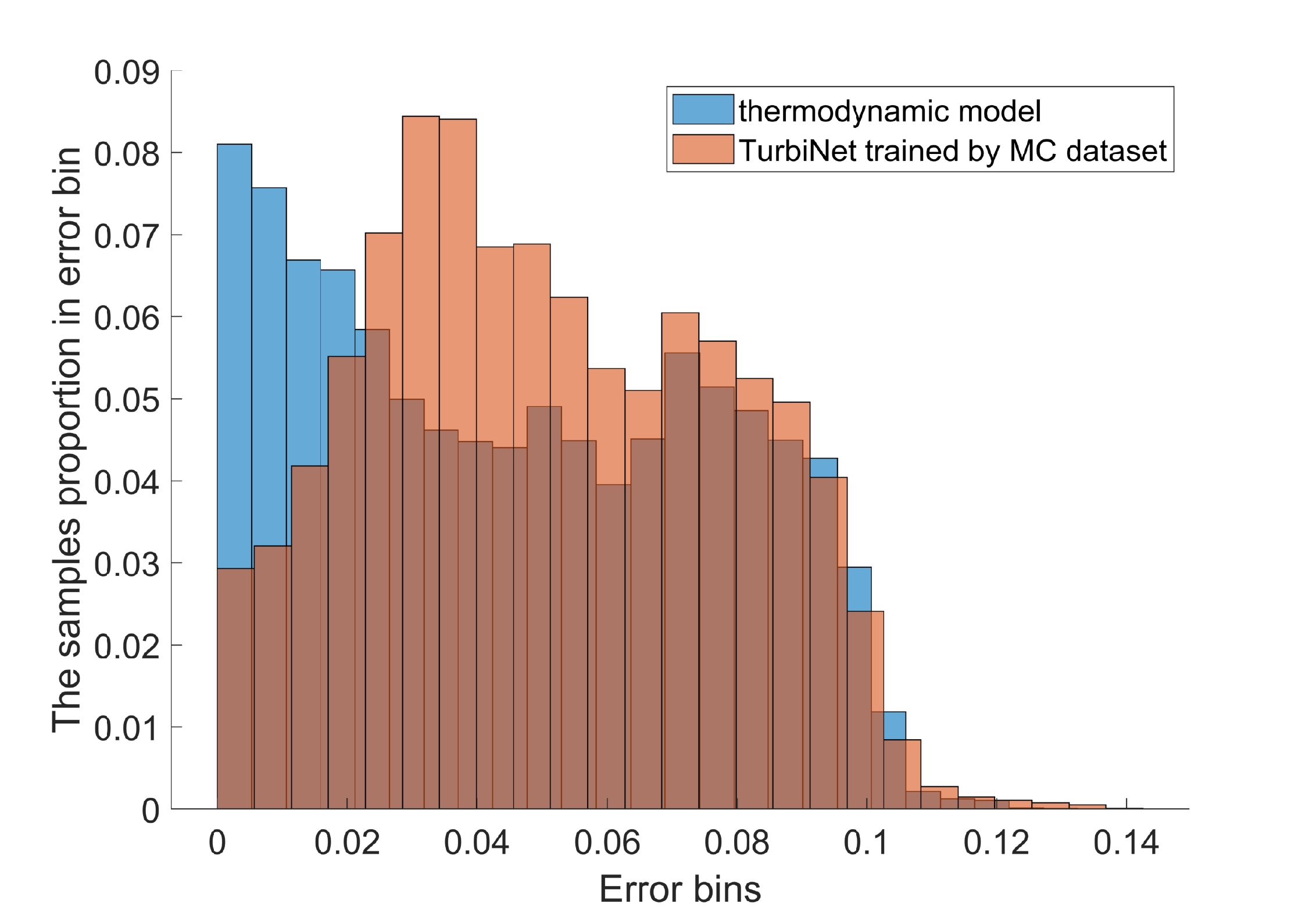}}
	\caption{T6 errors comparison between thermodynamic model and our network trained by MC dataset}
	\label{fig:T6errorsThermoTurb}
\end{figure}

We compare the T6 errors in flight testing dataset of our network trained only by Monte Carlo dataset with map fitting based thermodynamic model used in PROOSIS \cite{alexiou2007advanced, bala2007proosis}. The results are shown in Fig. \ref{fig:T6errorsThermoTurb}. Because our network is only train by Monte Carlo dataset generated by thermodynamic model and not trained by flight dataset, the accuracy of our model cannot be better. But from Fig. \ref{fig:T6errorsThermoTurb}, we can see that the error distribution of thermodynamic model and our network only trained by Monte Carlo dataset is similar. This indicates that our network pre-trained by Monte Carlo dataset  has the ability to reconstruct the component characteristics with the help of thermodynamic model.

\begin{figure}[htb]
	\centerline{\includegraphics[width=.9\columnwidth]{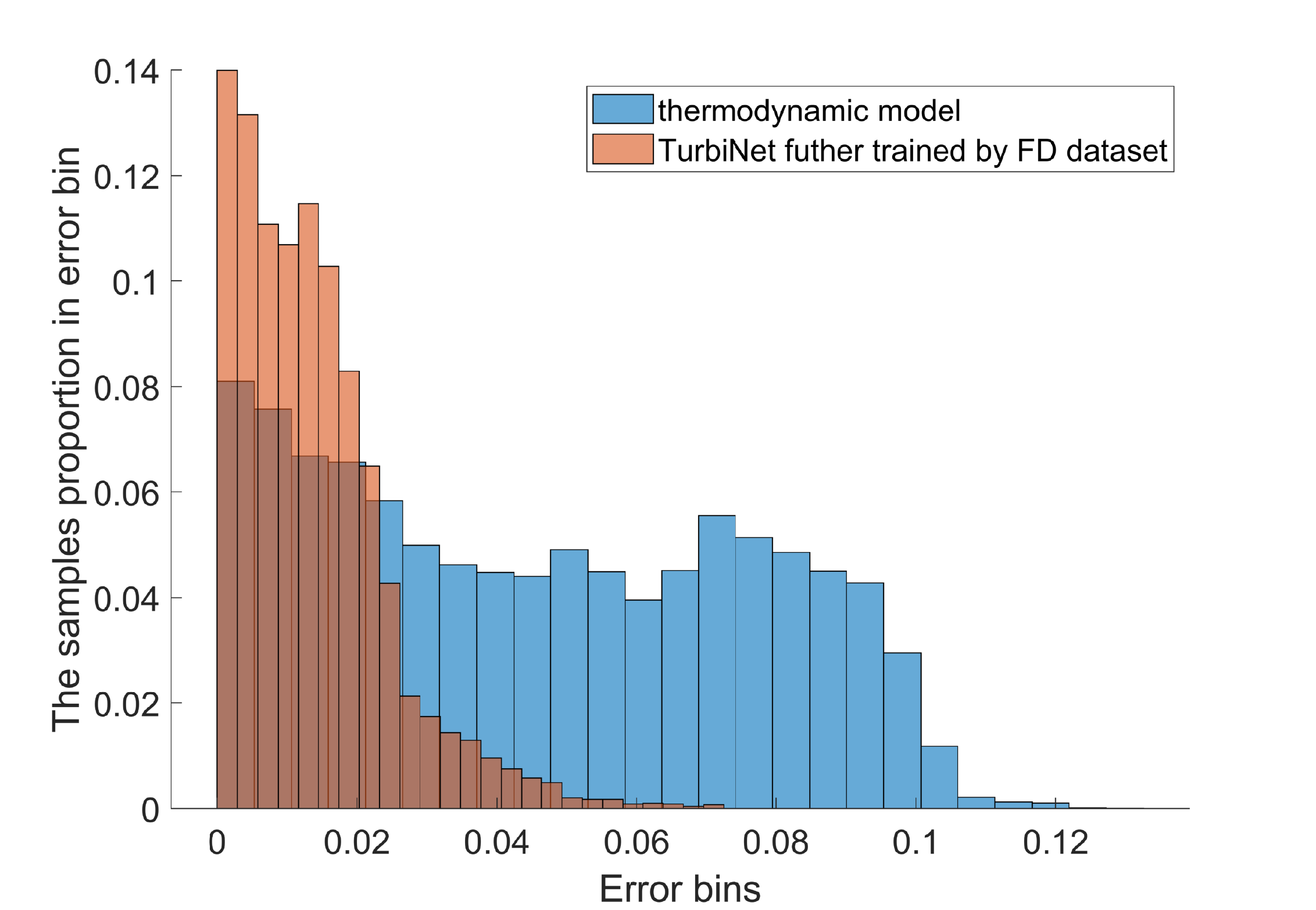}}
	\caption{T6 errors comparison between thermodynamic model and our network further trained by FD dataset}
	\label{fig:T6errorsThermoTurbFD}
\end{figure}

Furthermore, We compare our network further trained by flight training dataset with thermodynamic model. The results are shown in Fig. \ref{fig:T6errorsThermoTurbFD}. We can see that the max error of our network reaches 7.1\%, 5.2\% lower than thermodynamic model. This indicates that our network can extract the engine's individual characteristics and degeneration trend from flight data in the flight data training phase to evaluate the stations parameters more accurately.

\subsection{Comparisons with Neural Network Model}
\begin{figure}[htb]
	\centerline{\includegraphics[width=.9\columnwidth]{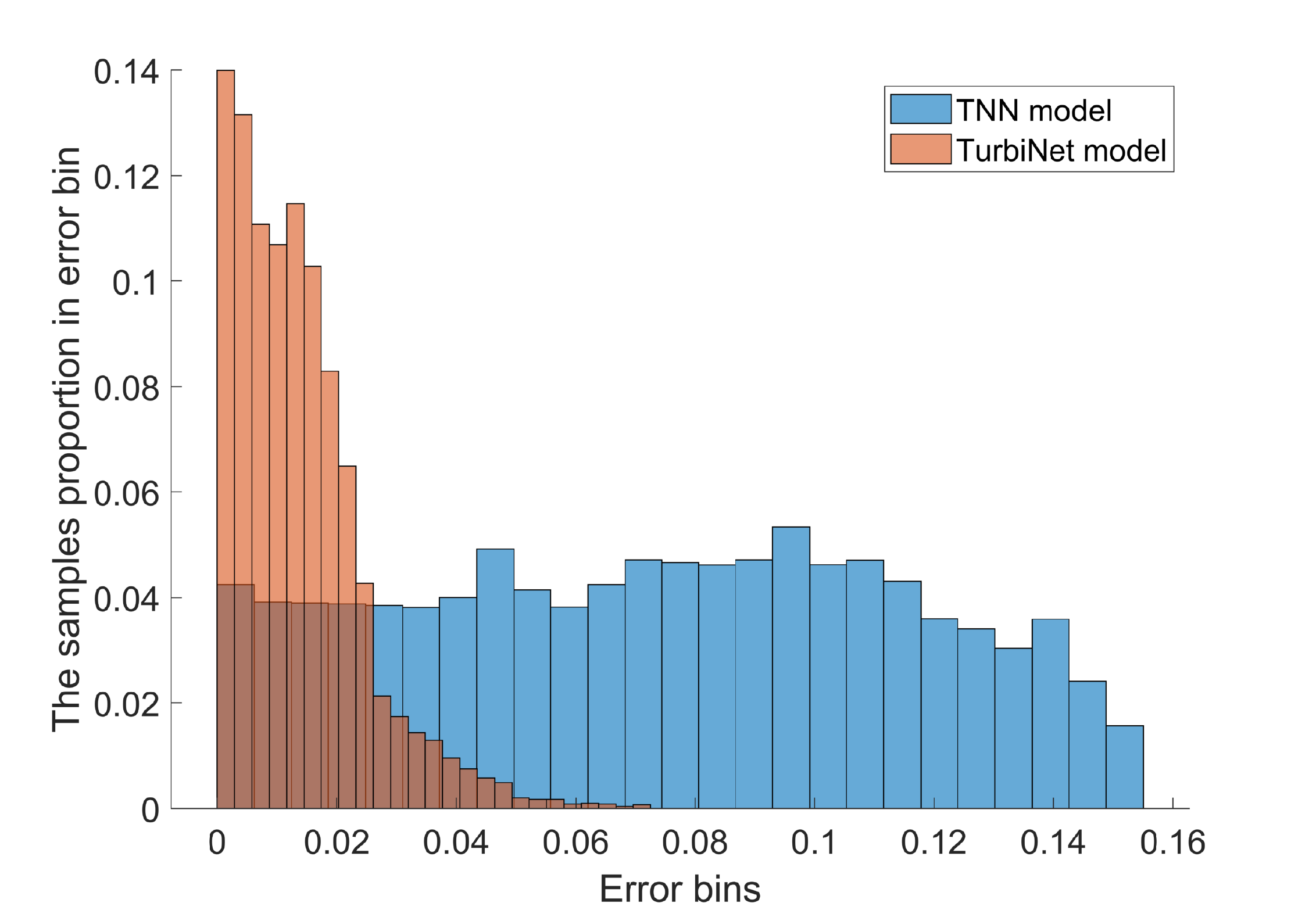}}
	\caption{T6 errors comparison between TNN model and our network}
	\label{fig:TNN_turbinet}
\end{figure}

To compare our network with pure neural network model, a neural network (denoted as TNN) of equivalent model volume is constructed to predict T6 in flight data. The neural network has $28$ hidden layers with the same number of hidden dimension as our network. The inputs are the same as our network's and the output is T6. The definition of the training loss is defined as Eq. \ref{eq:T6loss}. Because no other stations parameters can be evaluated by TNN, the thermodynamic error loss Eq. \ref{eq:Weqs} and \ref{eq:powereqs} cannot used as loss function. TNN is trained and tested using the same flight dataset under the same training configurations as our network. The comparison between our network and TNN is shown in Fig. \ref{fig:TNN_turbinet}. TNN's max T6 relative error reaches 15.8\%. The accuracy of TNN is too low to evaluate the performance of the engine, which verifies the rationality of our network's thermodynamic constrains and two-phase training process from another perspective.

\subsection{Accuracy comparisons of different methods}
The comparisons of different methods are shown in Tab. \ref{tab:comparison}. The accuracy is measured by the relative error (RE) between $T6$ in flight data and its evaluation $\hat{T6}$ by different methods:
\begin{equation}
error=\frac{|\hat{T6}-T6|}{T6}.
\label{eq:maxerror}
\end{equation}

\begin{table}[htb]
	\centering
	\caption{Accuracy comparison of different methods}
	\label{tab:comparison}
	\begin{threeparttable}
		\begin{tabular}{lllllll}
			\hline
			methods & mean & std & 25\% & 50\% & 75\% & max \\
			\hline
			PROOSIS & 0.046 & 0.030 & 0.018 & 0.044 & 0.072 & 0.123 \\
			TNN & 0.087 & 0.048 & 0.049 & 0.086 & 0.126 & 0.158 \\
			our network & 0.014 & 0.011 & 0.005 & 0.012 & 0.019 & 0.071 \\
			our network$\dagger$$^{\rm a}$ & 0.014 & 0.011 & 0.006 & 0.012 & 0.019 & 0.071 \\
			\hline
		\end{tabular}
		\begin{tablenotes}
			\footnotesize 
			\item[$^{\rm a}$] our network$\dagger$ is the iteration version described in \ref{sec:iter}.
		\end{tablenotes} 
	\end{threeparttable}
\end{table}

Among the table head, 'mean' is the mean RE of all testing samples, 'std' is the standard deviation of RE, '25\%' is the top '25\%' of RE, and 'max' is the max RE of all testing samples. We can see our network model gets the best accuracy in all measurement. In the 'max' measurement, our network is about 5\% higher than thermodynamic based model calculated in PROOSIS and 8\% higher than the pure data driven network with similar volume. The iteration version described in \ref{sec:iter} gets a similar accuracy with plain our network described in \ref{sec:fdtrain}, verifying the effectiveness of the iteration method.

\section{Conclusion}
In this paper, we propose a thermodynamic based and data driven hybrid network for gas turbine modeling. All components are modeled as neural network and their characteristics are extracted in a data driven way from Monte Carlo simulated data and flight data. Thermodynamic loss is introduced to ensure the training process converges to the turbine's dynamics and constrain the error transmission between component nets in hybrid network. The experiment shows that the accuracy of our hybrid network can reach about 7\% measured by max T6 relative error, 5\% better than map fitting based thermodynamic model and 8\% better than pure data driven method with similar model volume, verifying effectiveness of our proposed hybrid model.

\bibliographystyle{IEEEtran}
\bibliography{IEEEabrv, mybibfile}

%




\end{document}